# Ferroelectric domains in methylammonium lead iodide perovskite thin-films

Holger Röhm,[a†] Tobias Leonhard,[a†] Michael J. Hoffmann[b] and Alexander Colsmann[a*]

We explore the ferroic properties of methylammonium lead iodide perovskite solar cells by Piezoresponse Force Microscopy (PFM). In vertical and horizontal PFM imaging, we find domains of alternating polarization with a width of 90 nm which we identify as polarized ferroelectric domains. High-resolution photo-conductive atomic force micrographs under illumination also show alternating charge carrier extraction patterns which we attribute to the local vertical polarization components within the ferroelectric domains. The correlation of the sample properties with Atomic Force and Kelvin Probe Force Micrographs evidence the piezo-electric nature of the domains.

**Broader context:**
Within five years, methylammonium lead iodide ($MAPbI_3$) solar cells showed an unprecedented fast improvement towards power conversion efficiencies beyond 20%. However, arguably, the toxic and water-soluble lead compound may be an obstacle on their way to market. The quest for alternative, non-toxic photo harvesters is partly hampered by a lack of fundamental understanding of the $MAPbI_3$ properties and energy conversion mechanisms. As part of this process, the scientific community controversially discusses the importance of ferroic properties for the exceptional performance of $MAPbI_3$ light-harvesting layers, including claims of non-ferroelectricity, anti-ferroelectricity, ferroelectricity and ferroelasticity. Some simulations have predicted ferroelectricity in $MAPbI_3$ with alternating polarized domains ruling the charge carrier transport. Any experimental evidence towards ferroelectricity would therefore provide helpful guidance for the quest to find non-toxic $MAPbI_3$ replacements.

## Introduction

The unique material properties of Organometal Halide (OMH) Perovskites fostered an unprecedented fast improvement of power conversion efficiencies of Perovskite solar cells, recently surpassing 22% in lab-scale devices, and the increasing employment of OMH Perovskites in other optoelectronic applications.[1-4] Without the need for a p-n junction, heterojunction or other advanced device architectures that would require elaborated fabrication processes, this crystalline material exhibits long charge carrier lifetimes and charge carrier diffusion lengths along with efficient transport of both electrons and holes.[5,6] Instead, rather simple planar homo-junctions of OMH Perovskites and adjacent organic layers can be employed. Although, in the past few years, OMH Perovskites such as methylammonium lead iodide ($MAPbI_3$) sparked the interest of researchers all over the world, no conclusive concept of charge carrier transport in multi-crystalline thin-films has yet emerged. Using density functional theory (DFT) simulations, in early 2014, J. M. Frost et al. predicted ferroelectric properties of OMH Perovskites, originating from the orientation of the polar organic cations in the crystal lattice.[7] Assuming the existence of alternating ferroelectric domains in OMH Perovskite thin-films, they further suggested the formation of energetically favourable and hence separate pathways for electrons and holes at the interfaces of differently polarized domains, leading to reduced recombination losses. Shortly after, the first observation of ferroelectricity in small $MAPbI_3$ grains with typical sizes of 100 nm, each with unidirectional polarization, was reported by Y. Kutes et al.[8] According to simulations by S. Liu et al. as well as T. S. Sherkar and L. J. A. Koster, the formation of alternating domains of ferroelectric polarization within one grain would promote the formation of charge carrier transport paths along the domain interfaces.[9,10] Recently, the first observation of highly ordered ferroeleastic domains in OMH Perovskites was reported by Hermes et al.[11] Still, the ferroic properties are controversially discussed, including claims of non-ferroelectricity[12,13], anti-ferroelectricity[14], ferroelectricity[8] and ferroelasticity[11]. The variety of fabrication techniques for OMH Perovskite samples, including process additives and annealing procedures, often complicates the comparison of results. Additionally, different types of samples of the same material, such as powder samples, single crystals and thin films, can show different properties. For example, G et al.[13] reported a centrosymmetric (non-polar) crystal structure in pellet and single crystal samples of $MAPbI_3$, whereas Sewvandi et al.[14] concluded a non-centrosymmetric (polar) crystal structure in pellets of the same material.
In this work, employing Piezoresponse Force Microscopy (PFM), we find alternating polarized domains in $MAPbI_3$ thin-films as commonly incorporated in solar cells. $MAPbI_3$ was solution deposited using small amounts of methylammonium chloride, henceforth referred to as $MAPbI_3$(Cl). By correlating an AC voltage that is applied between the cantilever and the sample's bottom electrode, with the resulting local piezoelectric strain of the crystalline sample, PFM furthermore allows to distinguish the direction of domain polarization.[15] This polarization is correlated with photo-conductive Atomic Force Microscopy (pc-AFM), Kelvin probe force microscopy (KPFM) and the sample topography, letting us conclude that $MAPbI_3$(Cl) is indeed ferroelectric.

*a. Light Technology Institute, Karlsruhe Institute of Technology (KIT), Engesserstrasse 13, 76131 Karlsruhe, Germany*
*b. Institute for Applied Materials – Ceramic Materials and Technologies, Karlsruhe Institute of Technology (KIT), Haid- und-Neu-Strasse 7, 76131 Karlsruhe, Germany*
**. Contact: alexander.colsmann@kit.edu*
*†. The first two authors contributed equally to this work.*
*Electronic supplementary information (ESI) available. See DOI: XX.XXXX/XXXXXXX*

## Results & Discussion

External polarizability is one of the elementary properties of ferroelectric materials, yielding a hysteretic polarization response. However, the discordant reports in the literature on the observation of hysteresis loops or the lack thereof, to proof or disproof ferroelectricity in OMH perovskites underline the dilemma to measure such responses in this particular material. In accordance with previous reports in the literature[14,16] we observed a minor polarization hysteresis in $MAPbI_3$(Cl) samples upon applying a poling ramp between AFM tip and sample (-4 V to +4 V). However, after the measurement, the $MAPbI_3$(Cl) layer showed significant local burn-in damage. Notably, $MAPbI_3$(Cl) exhibits a high conductivity in comparison to most common ferroelectric ceramics such as $BaTiO_3$ or $Bi_4Ti_3O_{12}$ which are insulators. Due to its conductivity, any bias high enough to switch ferroelectric polarization in $MAPbI_3$(Cl) yields high current densities and may therefore damage the sample. Higher poling frequencies may reduce the sample damage, however, concomitantly, it also may result in incomplete poling.[14] For example, the small polarization retention previously reported by Coll et al. may stem from such incomplete poling of small material volumes close to the AFM tip. Therefore, poling experiments on $MAPbI_3$(Cl) can hint towards ferroelectric material properties, but the measurement results have to be interpreted with care.

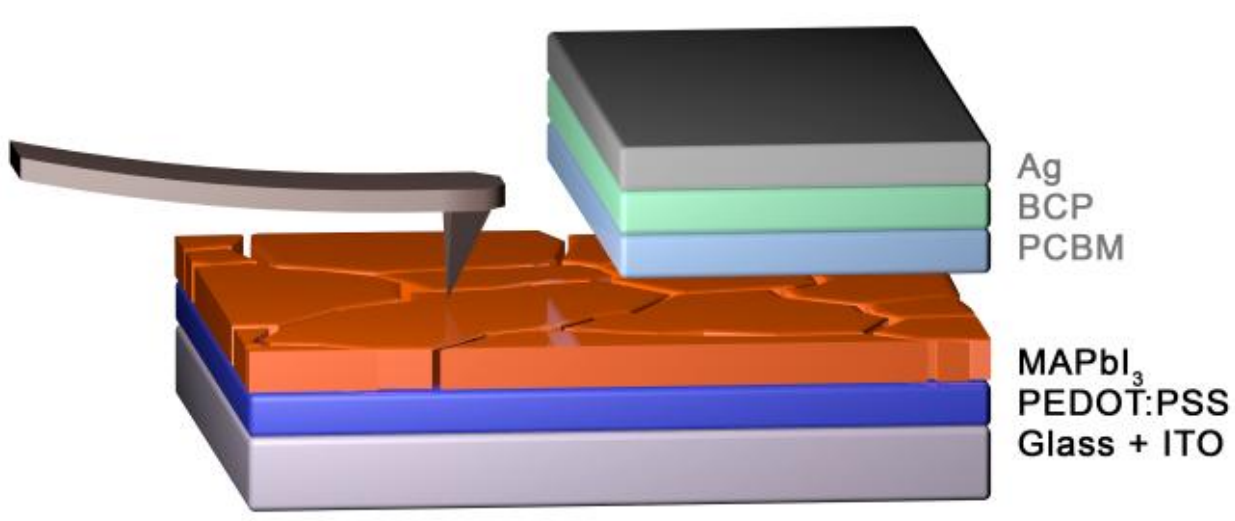

**Fig. 1** Scheme of the $MAPbI_3$ solar cell architecture and the sample characterization setup used to probe PFM, c-AFM, KPFM and the sample topography.

In contrast to the somewhat unreliable poling experiments on $MAPbI_3$(Cl), PFM unambiguously responds to a local inverse piezoelectric effect in the crystal.
In particular, ferroelectric versus ferroelastic properties in $MAPbI_3$ are controversially discussed. Ferroelasticity describes the relationship between external stress and internal strain in a crystal and the resulting local change of the crystal phase. Hence ferroelasticity can be probed by probing the change of internal polarization upon external mechanical deformation. Ferroelastic materials, as reported earlier,[11] can form twin-domains when subjected to external mechanical stress. However, these domains cannot be probed by PFM, since spatial-inversion symmetry is not influenced by ferroelastic distortion in a crystal.[17] In contrast, ferroelectricity correlates the change of polarization in a crystal with spontaneous or external polarization. Therefore, ferroelectric crystals expand or contract in response to a local external electrical field. This so-called inverse piezoelectric effect can be probed by PFM.

To investigate the piezo-responsivity of $MAPbI_3$(Cl), we performed PFM measurements on about 300 nm thick Perovskite layers that enable power conversion efficiencies of up to 14% in low-hysteresis solar cells (Supplementary Information, Fig. S1), when included in a planar ITO/PEDOT:PSS/$MAPbI_3$(Cl)/$PC_{71}BM$/BCP/Ag device architecture (Fig. 1).
For PFM sample preparation, we followed the solar cell fabrication process all to the deposition of the Perovskite layer, omitting the deposition of the electron transport layers and the metallic top electrode. Employing solvent-annealing, we achieved layers comprising flat, featureless and large Perovskite grains with average diameters of several micrometers. The flat grain surface is important to minimize cross-talk from the sample topography to the highly sensitive piezoelectric response of the sample.
Importantly, all AFM images were recorded in a glovebox under nitrogen atmosphere, since any adventitious surface contamination, *e.g.* with water, can screen the properties of the bulk below. The AC frequency applied to the cantilever during PFM measurements was deliberately chosen close to the resonance frequency of the AFM cantilever in contact with the sample to enhance the signal intensity. However, using an AC frequency close to the resonance frequency comes at cost of adding noise to the phase signal of the lock-in amplifier, therefore impeding the correlation of signal and domain orientation.[15] Since non-enhanced PFM did not yield reliable readouts for various different types of cantilevers, we conclude the piezoelectric coefficient of $MAPbI_3$(Cl) to be rather small.

Patterns of polarized domains within one $MAPbI_3$(Cl) grain become visible in the PFM images in Fig. 2a, featuring parallel stripes of polarization with an average width of about 90 nm. As known from ferroelectric materials such as $BaTiO_3$ or $(K_{0.5}Na_{0.5})NbO_3$, such orientation patterns can self-organize by spontaneous polarization to reduce the mechanical strain within grains during phase transformation from cubic to tetragonal or orthorhombic phase.[18,19] Resulting depolarizing fields render alternating domains energetically more favourable than large crystal domains of the same polarization that would lead to an overall polarization of the whole grain.

The alternating domains visible in PFM only allow for a ferroelectric or anti-ferroelectric nature of non-annealed $MAPbI_3(Cl)$ and a ferroelectric nature after annealing, which is in accordance to the findings of Sewvandi et al. in compressed $MAPbI_3$ pellets.[14]

We note that a non-centrosymmetric piezoelectric material that is not ferroelectric, cannot form alternating polarized domains within a monocrystalline grain. Such domains of the same polarization would - by definition - not be part of the same single crystal in a purely piezoelectric compound.

We therefore conclude that $MAPbI_3(Cl)$ is indeed ferroelectric. Here we observed ferroelectric domains of different orientations formed by spontaneous polarization within the crystal. The domain patterns are visible in both vertical and horizontal PFM images, indicating that vertical and in-plane components contribute to the ferroelectric polarization (Supplementary Information, Fig. S2a and b).

We observed similar linear domain structures in thermally annealed samples that were not subdued to solvent-vapor, however, the less homogeneous and smaller grains added noise to the measurement (Supplementary Information, Fig. S3).

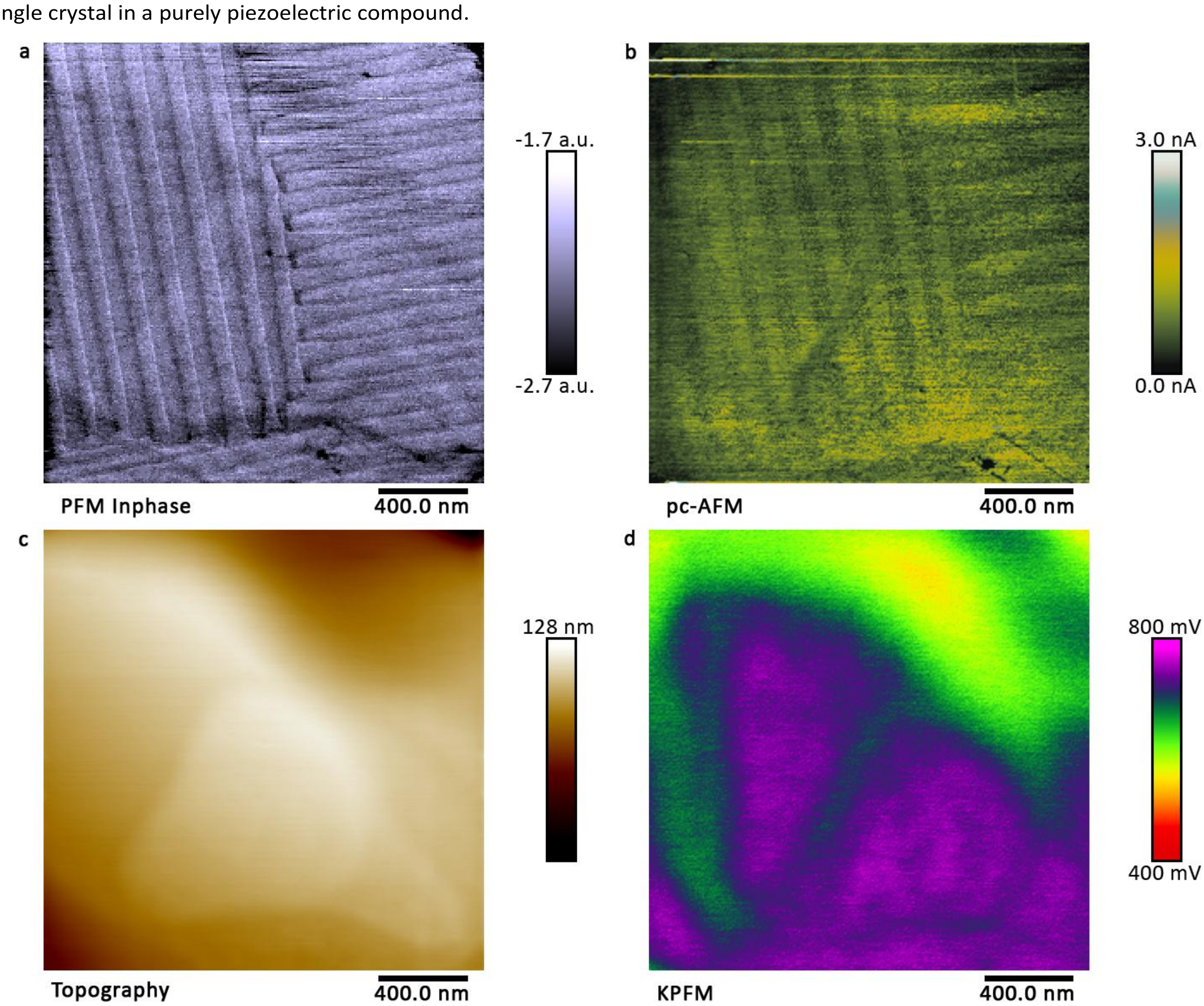

**Fig. 2** High-resolution (a) PFM, (b) pc-AFM, (c) topography and (d) KPFM images of a typical $MAPbI_3(Cl)$ sample surface (further examples are provided in the Supplementary Information, Fig. S2, S3, S5, S6, S7, S8, S9 and S10). All four images in this graph show the surface of the same grain. PFM imaging reveals patterns of highly ordered ferroelectric domains of different polarization. The same patterns are visible in pc-AFM mode. For reference, the flat and featureless shape of the grain as visible in the topography excludes any relevant influences of the topography on the PFM or pc-AFM measurements. Neither did the KPFM image show any influence of the ferroelectric polarization on the surface potential. The work function variations of up to 300 mV between individual grains may originate from crystal face variations.

Such polarized domains are also visible in other $MAPbI_3$ thin-films as exemplified in the Supplementary Information (Fig. S9) on layers deposited without methylammonium chloride. Although the omission of methylammonium chloride changes the processing conditions and hence yields smaller grains and rougher layers, alternating polarized domains become clearly visible.

We note that the influence of the dipolar organic cation on the ferroic nature of $MAPbI_3$(Cl) is still unclear, although a certain mutual influence between cation rotation and distortion of the central $PbI_6$ octahedron in the lattice seems likely.[20]

Occasionally, we observed a 90° continuation of the polarization pattern within one grain (Fig. 3a+b). From this 90° continuation of some ferroelectric domains, we further conclude a horizontal rotation of the ferroelectric polarization orientation versus the orientation of the domain itself. For reasons of symmetry, we conclude a rotation angle close to 45°. At grain boundaries, the ferroelectric domains can terminate with reduced or expanded width and often continue along falling grain flanks.

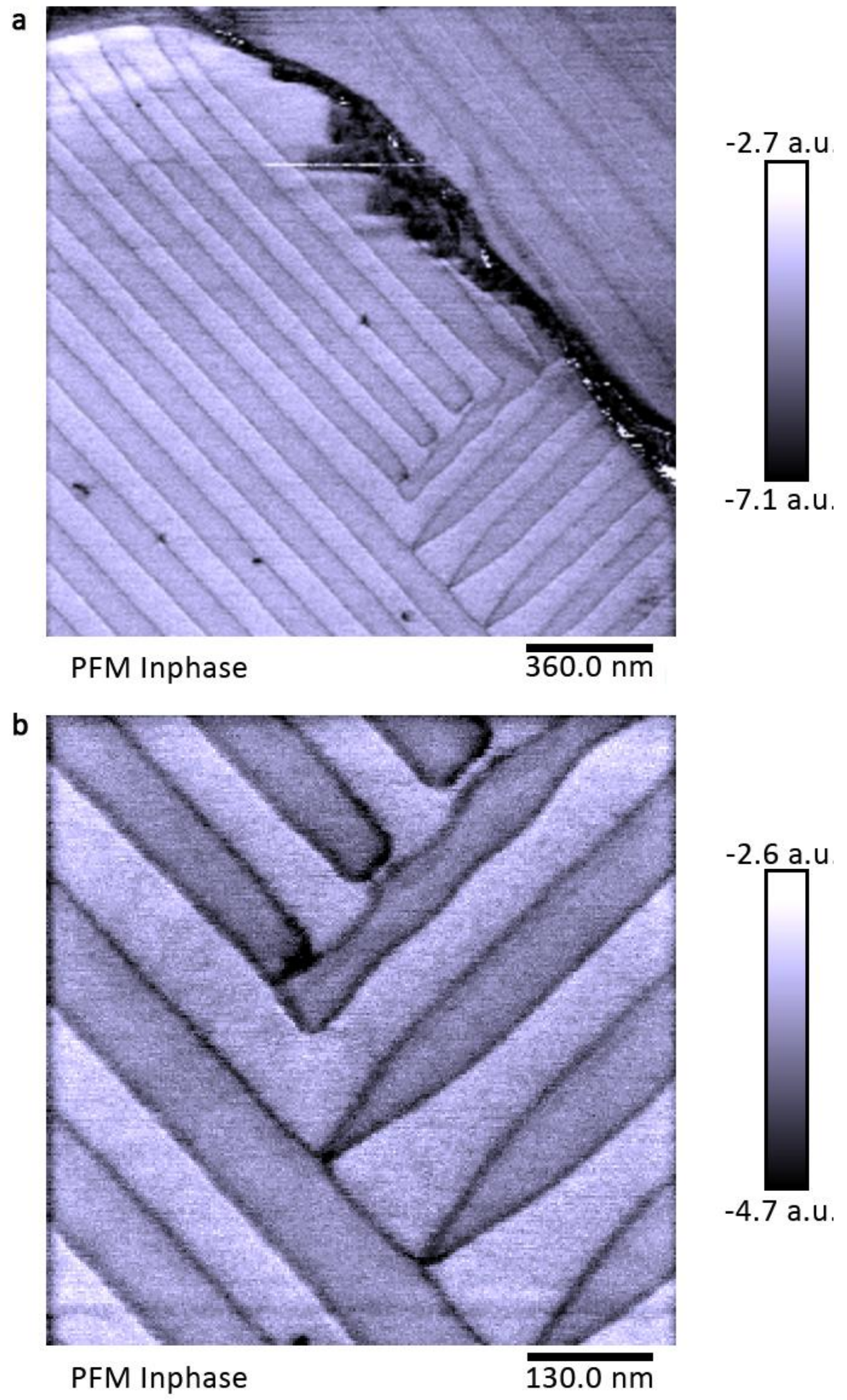

**Fig. 3** (a) High-resolution PFM image of a ferroelectric $MAPbI_3$(Cl) domain pattern. The grain boundary is visible as dark line across the image. The grain flank in the top left corner reveals vertically continuing domains within the grain. (b) Detailed PFM image of a smaller area of the same sample, showing a 90° continuation of the polarized domain.

In contrast to previous reports on externally polarized OMH Perovskite layers, we found the self-organized line-shaped ferroelectric domains to be stable at room temperature.[16] When repeating PFM measurements after several hours on the very same grains, we observed no change in the domain shapes or contrast. Polarized domains can still be observed after more than two months of sample storage under dry nitrogen atmosphere. $MAPbI_3$(Cl) layers comprising small grains of similar size on the order of the average ferroelectric domain size, as investigated by Y. Kutes et al., may not exhibit domain interfaces within grains, but only at grain boundaries.[8] Ferroelectric materials indicate a critical grain size below which only single-domain grains exist.[21] And even if alternating domains exist in small OMH Perovskite grains, cross-talk from their rough topography may well obscure any ferroelectric domains.

Assuming that the interfaces of domains with different polarization occur mainly along the main crystal faces, the domain patterns also provide information on the grain orientation.[22] Since mostly 0° and 90° angles of domain interfaces occur within one (large and flat) grain, we conclude a correlation between grain orientation of the tetragonal $MAPbI_3$(Cl) crystal and the flat interface underneath.

The ferroelectric domains probed by PFM consistently extend over grain flanks, indicating that the domains are indeed a bulk property and continue throughout the whole grain. This finding may be of particular importance for the vertical charge carrier transport properties of the grains, since continuous domain boundaries can provide pathways to the top or bottom layer interfaces.

To investigate the influence of ferroelectric domains on the extraction of photo-generated charge carriers from the $MAPbI_3$(Cl) bulk, we employed photo-conductive AFM (pc-AFM). Although we operated pc-AFM imaging close to its spatial and photo-current resolution limits, the striped patterns of the ferroelectric $MAPbI_3$(Cl) domains became visible in the pc-AFM image in Fig. 2b. On every other domain, we found an about 25% enhanced extraction of charge carriers that may originate from a polarization component of the $MAPbI_3$(Cl) crystal in vertical direction. In absence of a top electrode, photo-generated charges populate the crystal surface. The charge carrier collection at the AFM tip is enhanced if a favourable ferroelectric polarization causes an additional electric field between the AFM tip and the $MAPbI_3$(Cl) bulk. Importantly, the ferroelectric domains are visible in the pc-AFM images even if no PFM measurements were performed before. Hence, the low AC voltage that is used in PFM mode, does not affect the polarization patterns. We note that, on some grains, the variations in current amplitude are obscured by noise. The spatial resolution of pc-AFM measurements is limited by the tip size (here: $r_{tip}$=20 nm) and tip geometry and requires very good contact between sample surface and probing tip, rendering it highly sensitive to variations thereof. Hence, the flat grain surfaces in this work are an important asset to the recording of meaningful pc-AFM images.

In order to exclude any measurement errors stemming from cross-talk of the grain topography or from surface contamination, we further performed topography and Kelvin Probe Force Microscopy (KPFM) measurements (Fig. 2c and d) on the very same spots as the PFM and pc-AFM measurements. Since the grain's surface topography does not exhibit any striped patterns as found in the PFM or pc-AFM images, we exclude any influence of the topography on the domain patterns in Fig. 2a. The KPFM image in Fig. 2d shows a homogeneous work function on flat grain surfaces, again without any contributions from the domains. The vertical polarization component perpendicular to the surface that we concluded from pc-AFM measurements, do not affect the KPFM signal. Rounded surfaces and angled crystal flanks result in a change of the surface potential as the face of the crystal deviates from the three main faces of the perovskite. This observation is in line with KPFM measurements of $MAPbI_3(Cl)$ layers comprising only small grains with an average diameter of about 300 nm, where we found significant work function variations between the surfaces of different grains (Supplementary Information, Fig. S4) that we attribute to strong variations in the small grains' crystal orientation. We note that the effect of varying Perovskite surface work function is of particular importance in all optoelectronic Perovskite devices such as perovskite LEDs and solar cells, since the surface work function determines band bending and therefore charge carrier transport at interfaces.[23] Besides crystal orientation, KPFM is highly sensitive to variations of material composition.[24,25] Therefore, the low variation of <100mV between flat grain surfaces also rules out any noteworthy surface contamination by precursor remnants such as lead iodide ($PbI_2$), methylammonium iodide (MAI) or methylammonium chloride (MACl).

All effects described herein have been observed on many samples, with Fig. 2 being only one example. Further images are attached as part of the Supplementary Information (Fig. S2, S3, S5, S6, S7, S8, S9, S10 and S12). The same principle that we show here on a methylammonium (MA) based Perovskite, is likely to be applicable in formamidinium (FA) Perovskites due to the similar, yet weaker dipolar character of FA.[26] FA or mixtures of FA and MA are employed in OMH Perovskite solar cells surpassing PCEs of 20%.[1]

Ordered domains in the direction of charge carrier extraction as observed here may be beneficial for low-loss charge carrier transport in solar cells. However, OMH Perovskites may exhibit similar properties in disordered domain boundaries in very small grains or grains of high crystal defect density that do not allow for highly ordered domain structures as already discussed by Frost et al.[7] While the actual level of influence of polarized ferroelectric domains on the device performance is yet unclear, previously reported simulations suggest a significant positive or negative impact on charge carrier transport depending on the nanostructure of said domains.[7,9,10]

## Conclusions

In conclusion, we observed self-organized, room temperature-stable ferroelectric domains in OMH Perovskites and correlate PFM, pc-AFM, AFM and KPFM measurements of the same grains in Perovskite layers that are commonly used in efficient solar cells. These domains do not affect topography or surface work function but influence charge carrier extraction at the layer surface. The distinct domain patterns visible in both vertical and horizontal PFM images can only originate from ferroelectricity in $MAPbI_3(Cl)$. These results give new insight into the material properties of OMH Perovskites and pave the way for further device optimization. The knowledge of the ferroelectric nature of $MAPbI_3(Cl)$ as employed in the currently most efficient Perovskite solar cells is a valuable asset for the smart design of new (lead-free) materials. Exploration of new crystalline materials that exhibit the advantageous properties of OMH Perovskites, but avoid their instability and toxicity issues, is of paramount interest to researchers and component manufacturers alike, since those disadvantages have become evident to be the main obstacles on the way to commercialization of OMH Perovskite based devices.

## Acknowledgements

We thank S. Gärtner, V. Meded, S. Wagner, and H. Kalt for fruitful discussions. J. Fragoso laser-structured the ITO electrode. This work was funded by the Baden-Württemberg Foundation (project NanoSolar, contract no. CT-9). The AFM was made available through funding by the Federal Ministry of Education and Research under contract 03EK3504 (project TAURUS). The authors thank the DFG Center for Functional Nanostructures (CFN) for support.

## Methods

*Device fabrication*: Poly(3,4-ethylenedioxythiophene):poly(4-styrenesulphonate) (PEDOT:PSS, VPAI 4083, Heraeus Clevios) was diluted with isopropyl alcohol (IPA, 1:3 v/v) and blade-coated on cleaned indium tin oxide (ITO) glass substrates (4 mm/s, gap 400 µm, platen T=65 °C). The films were then annealed (140 °C, 20 min) to remove water residues, yielding a layer thickness of 25 nm. Lead iodide ($PbI_2$, Sigma-Aldrich) was dissolved in dimethyl sulfoxide (DMSO) and subsequently filtered with a hydrophilic membrane filter (pore size 0.25 µm). Then the solution was mixed with 1-chloronaphtalene (CN, 100:1 v/v) and blade-coated atop PEDOT:PSS (3 mm/s, gap 70 µm, T=60 °C). After 20 s, the films were dried under nitrogen flow to obtain a smooth and homogeneous $PbI_2$ layer. The substrates were transferred into a glovebox with nitrogen atmosphere. Methylammonium iodide (MAI) and methylammonium chloride (MACl) (Lumtec Ltd.) were mixed (9:1 w/w) and dissolved in IPA (40 mg/mL). 60 µL of this solution were spread over the $PbI_2$ layer. After 5 s of soaking, any solution excess was removed by spin-coating (3500 RPM). The substrates were exposed to dimethylformamide vapour (DMF, 60 min), after evaporating DMF (20 µL) under a petri dish on a hotplate (T=100 °C).[27] The films first changed colour to bright yellow and then turn dark brown within about 1 min, indicating the formation of a crystalline perovskite layer (thickness 300 nm). Subsequently, [6,6]-phenyl $C_{71}$-butyric acid methyl ester solution ($PC_{71}BM$, 40 µl, 20 mg/mL) was spin-coated from chlorobenzene solution (CB; 1000 RPM, 30 s; 4000 RPM, 10 s; layer thickness 20 nm). After sample annealing (100 °C, 20 min), bathocuproine (BCP, 60 µL, 0.5 mg/mL) was spin-coated from ethanol solution atop $PC_{71}BM$ (4000 RPM, 20 s. layer thickness 5 nm).[28] Finally, Ag was thermally evaporated through a shadow mask to define the top electrode ($10^{-6}$ mbar, layer thickness 100 nm).

*AFM measurements*: The AFM measurements were performed on an atomic force microscope (AFM, Dimension Icon, Bruker) under nitrogen atmosphere in a glovebox to protect the samples from degradation in humid air. Electrical conductive tips (Bruker AFM Probes) with Platinum-Iridium coated cantilevers were used for PFM and KPFM. A Platinum-Silicide cantilever was employed for pc-AFM measurements. Highest resolutions of ferroelectric domains were obtained with PFM in contact resonance using SCM-PIC tips possessing a free resonance frequency of 10 kHz and a spring constant of 0.1 N/m. The first resonance frequency for the tip-sample system was in the range of 30-40 kHz. The AC bias frequency applied to the sample was set at the lower rising flank of the resonance peak. All PFM data is depicted in the PFM-Inphase channel, combining the amplitude with the cosine of the phase θ (Inphase=Amplitude · cos θ).[15] For pc-AFM measurements, the sample was illuminated from below through a glass fibre connected to a solar simulator (Newport, Modell 67005, 75 W Xenon lamp)

Fig. 2a: 1500 mV AC bias, 34.75 kHz, vertical, scan angle 90°, probe SCM-PIC-V2. Fig. 2b: under illumination, +800 mV bias, probe SCM-PtSi. Fig. 2c+d: Drive amplitude 3000 mV, under illumination, probe SCM-PtSi.

Fig. 3a+b: 1000 mV AC bias, 34.43 kHz, vertical, scan angle 90°, probe SCM-PIC-V2.

*Optoelectronic characterization*: Current density-voltage (*J-V*) curves were measured using a Keithley 2400 source-meter unit in the dark or under illumination with an Oriel solar simulator (1000 W/m², AM 1.5), that was monitored with a calibrated Si-reference cell. Sweeps were conducted at a scanning speed of 100 mV/s in reverse (1.2 V → -0.2 V) and forward direction (-0.4 V → 1.2 V). The power conversion efficiency was determined from semi-steady state measurements at a scanning speed of 5 mV/s around the maximum power point.

# Supplementary Information

## Ferroelectric domains in methylammonium lead iodide perovskite thin-films

*Holger Röhm, Tobias Leonhard, Michael J. Hoffmann and Alexander Colsmann**

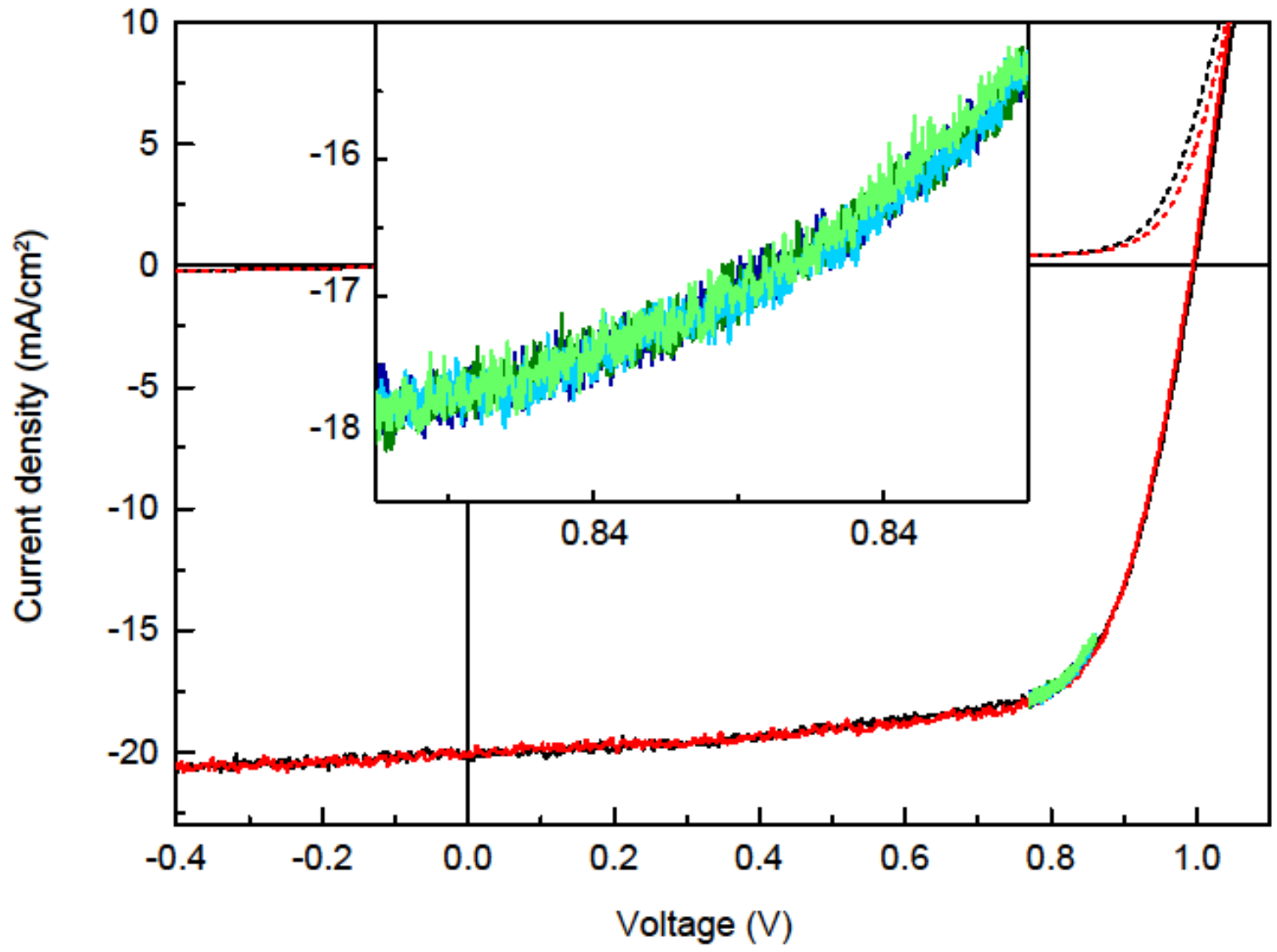

**Fig. S1**. Almost hysteresis-free *J-V* curves of a typical $MAPbI_3$(Cl) solar cell in forward (red curve) and reverse (black curve) direction (scanning speed 100mV/s). Inset: The PCE of 14% was derived from forward (dark and light blue) and reverse (dark and light green) semi-steady state measurements (5 mV/s) around the maximum power point. No external electrical poling was conducted prior to the measurement.

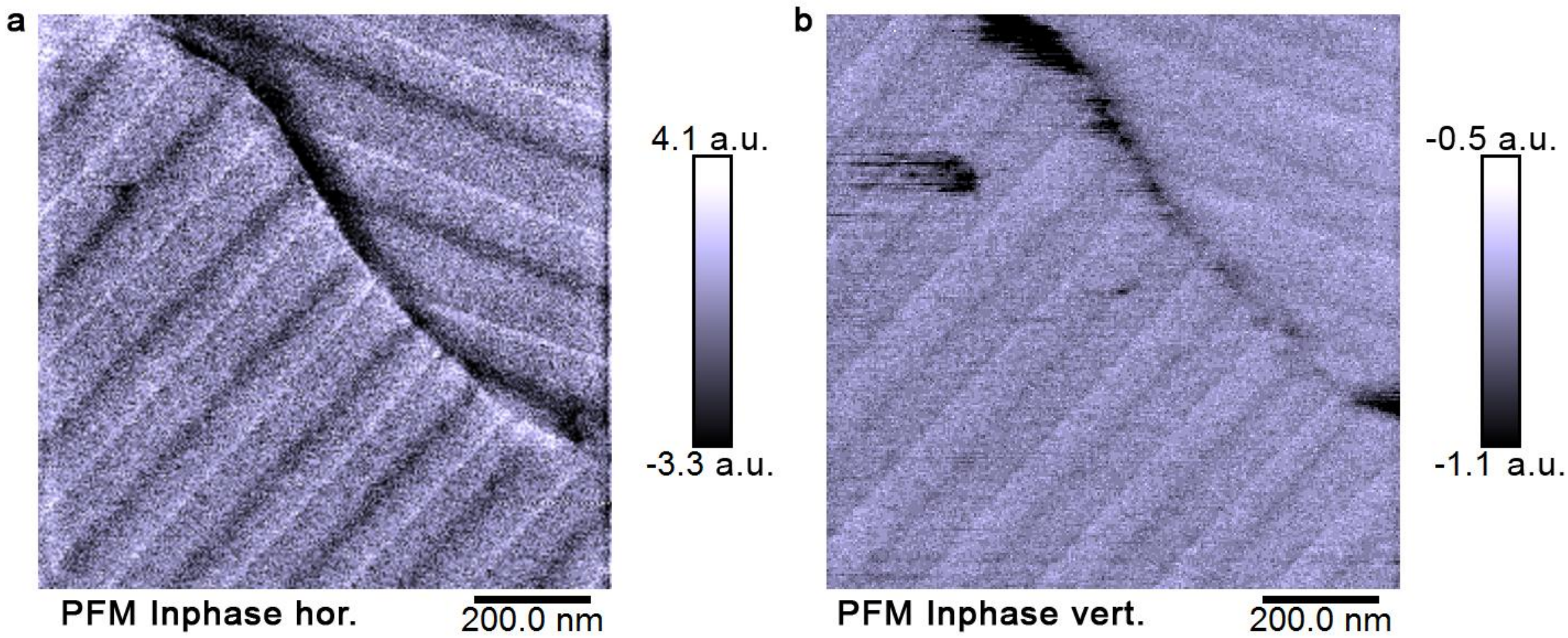

**Fig. S2**. Comparison of (a) horizontal (in-plane) and (b) vertical PFM signals of a $MAPbI_3(Cl)$ grain boundary. Both channels show the same domain features, whereas different cantilever stiffness for vertical and horizontal deflection hampers quantitative analysis of vertical and horizontal signal magnitudes, respectively. The vertical cantilever response can be influenced by in-plane polarization via buckling of the cantilever, but the horizontal response is decoupled from the vertical polarization. Both images shown here were recorded in direct succession using the same PFM settings.

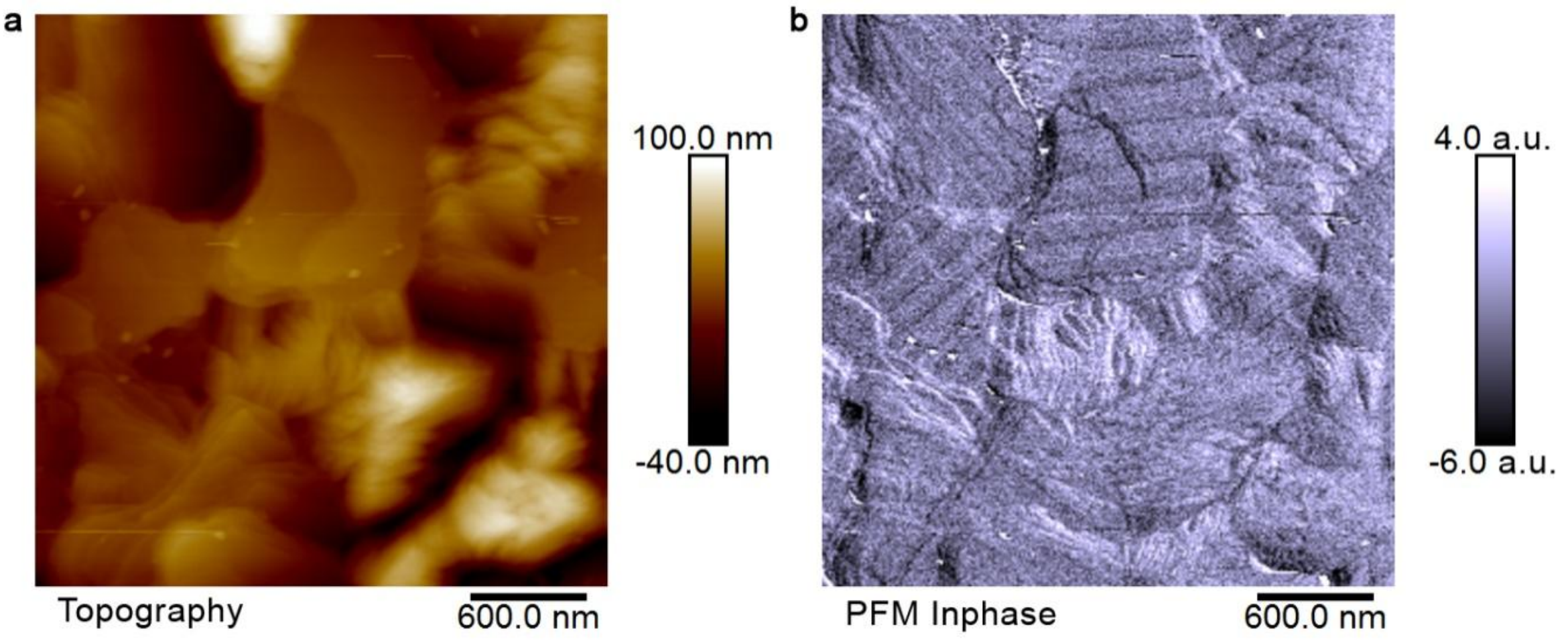

**Fig. S3**. (a) Topography and (b) PFM Inphase image of the same $MAPbI_3(Cl)$ grain after thermal annealing and without subduing the sample to solvent vapor. The flat grains show line-shaped domain structures.

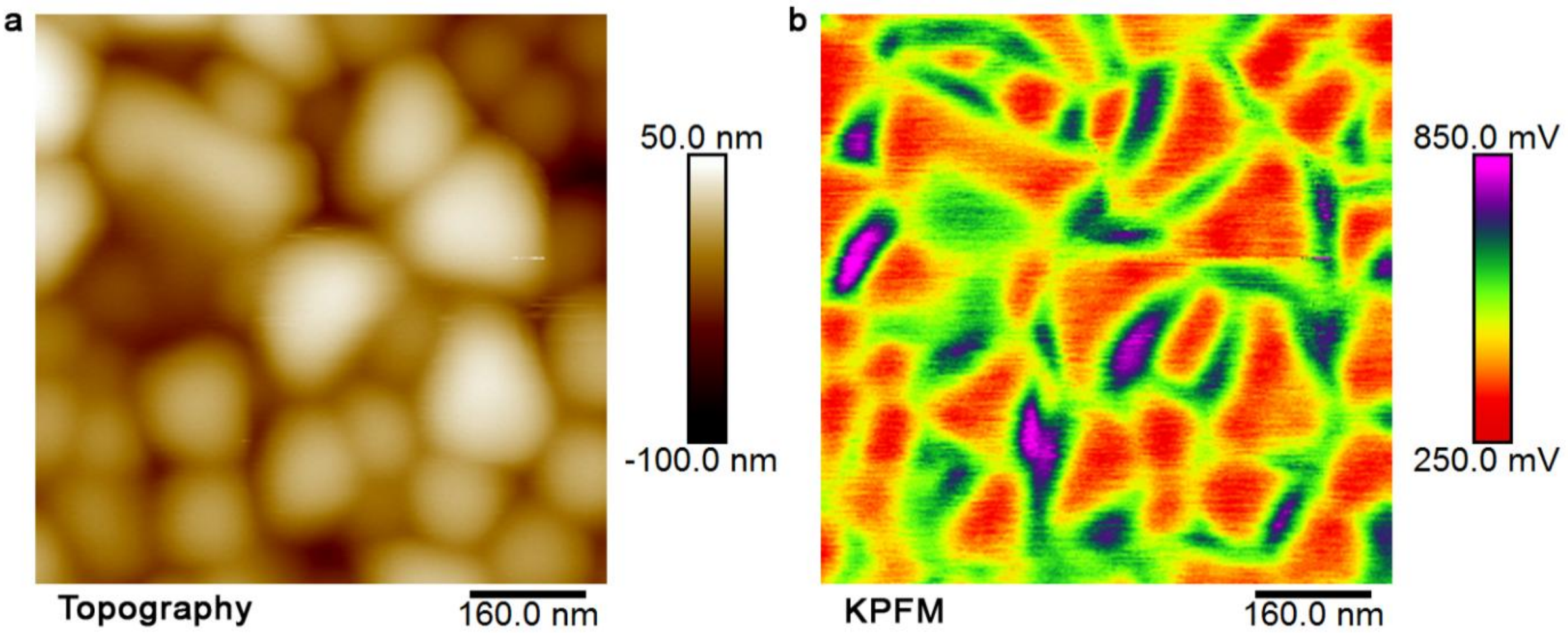

**Fig. S4**. (a) Topography and (b) KPFM image of the same $MAPbI_3(Cl)$ sample featuring small grains (prepared in air and after thermal annealing). Flanks of tilted cuboid grains exhibit high surface potential on flat grain flanks and low surface potential at rounded edges.

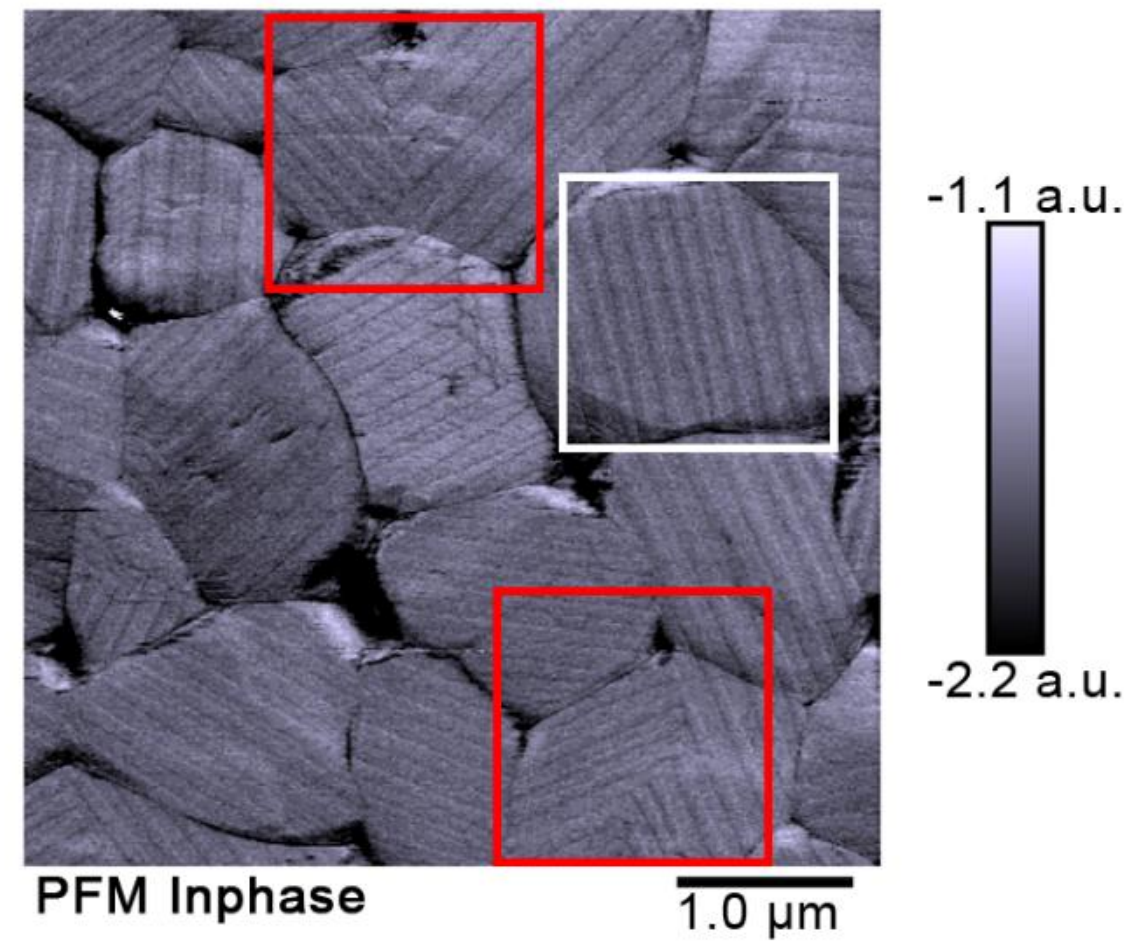

**Fig. S5**. PFM image of ferroelectric $MAPbI_3(Cl)$ domains on multiple grains show typical patterns of polarization: Some grains exhibit only one pattern of parallel domains (white square), whereas bigger grains often feature multiple areas of similar patterns, rotated approximately 90° relative to each other (red squares).

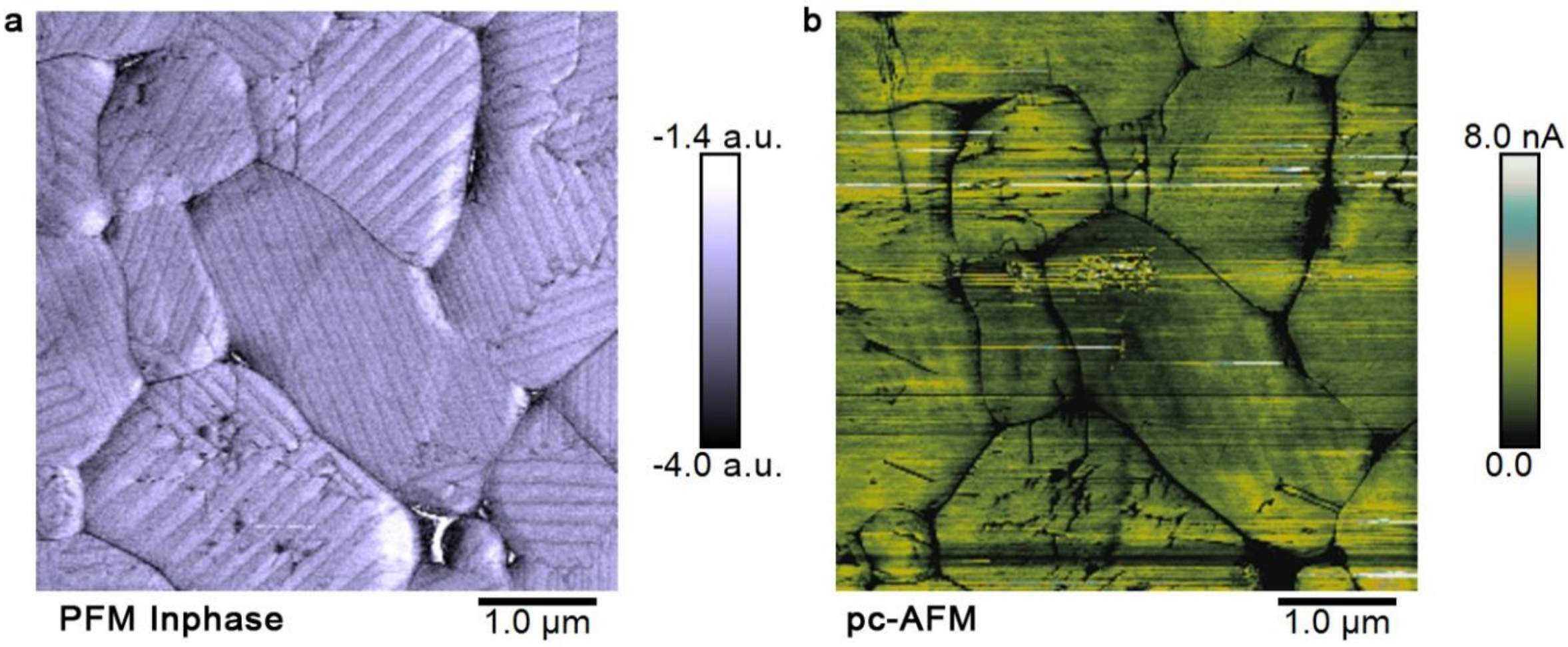

**Fig. S6**. (a) PFM image of Perovskite grains showing different ferroelectric domains and (b) pc-AFM image of the same grains showing the correlation of the local photo current with the ferroelectric domains.

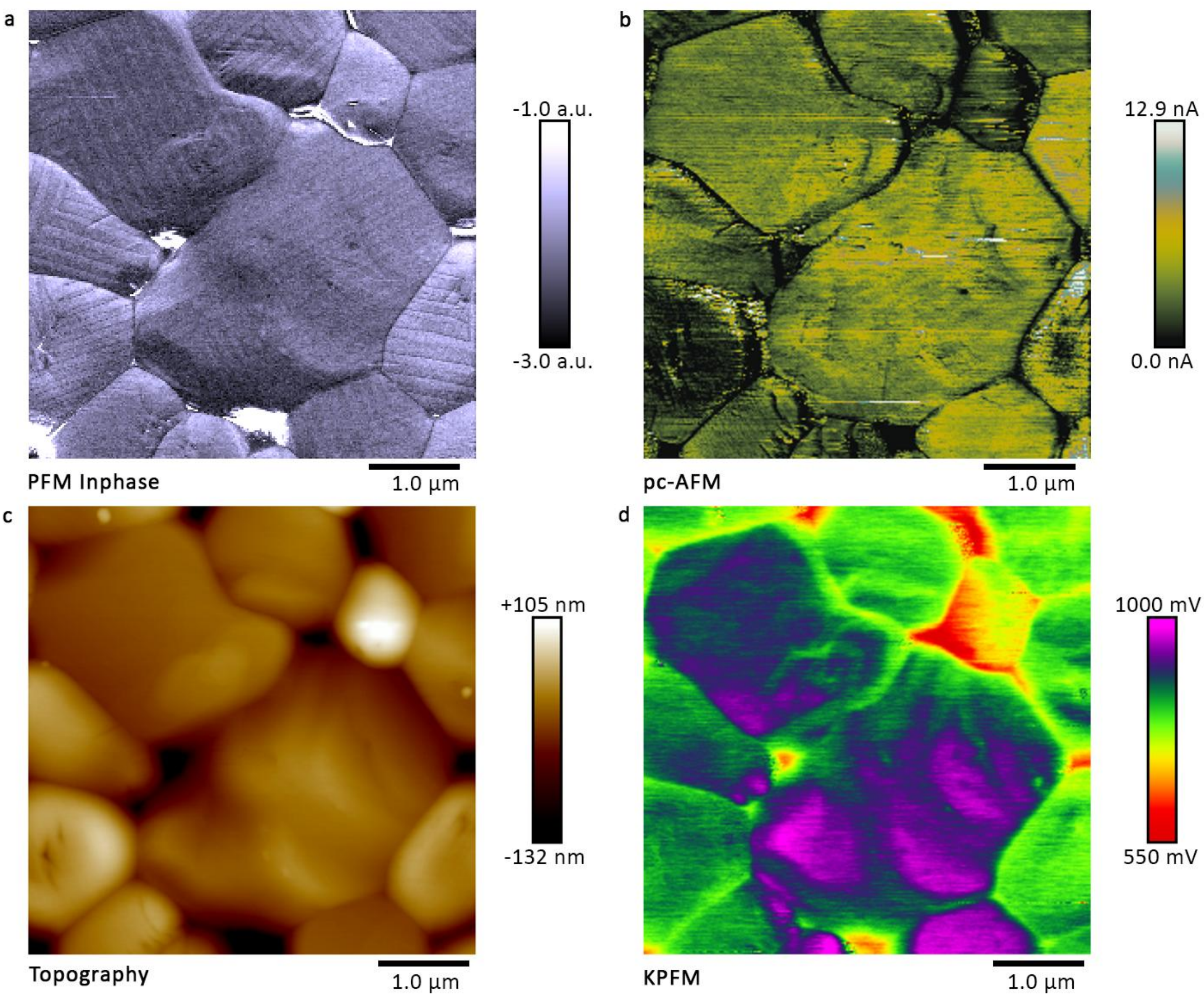

**Fig. S7**. (a) PFM, (b) pc-AFM, (c) topography and (d) KPFM images of the same flat $MAPbI_3$(Cl) grain. This measurement was performed on a different sample than depicted in Fig. 2 or Fig. S7, yet shows similar grain properties. Neither topography nor KPFM are influenced by the ferroelectric domain patterns visible in the PFM measurements. The pc-AFM image, measured under short-circuit conditions, exhibits the same striped patterns as the PFM image, indicating an influence of the ferroelectric polarization on the local charge carrier extraction.

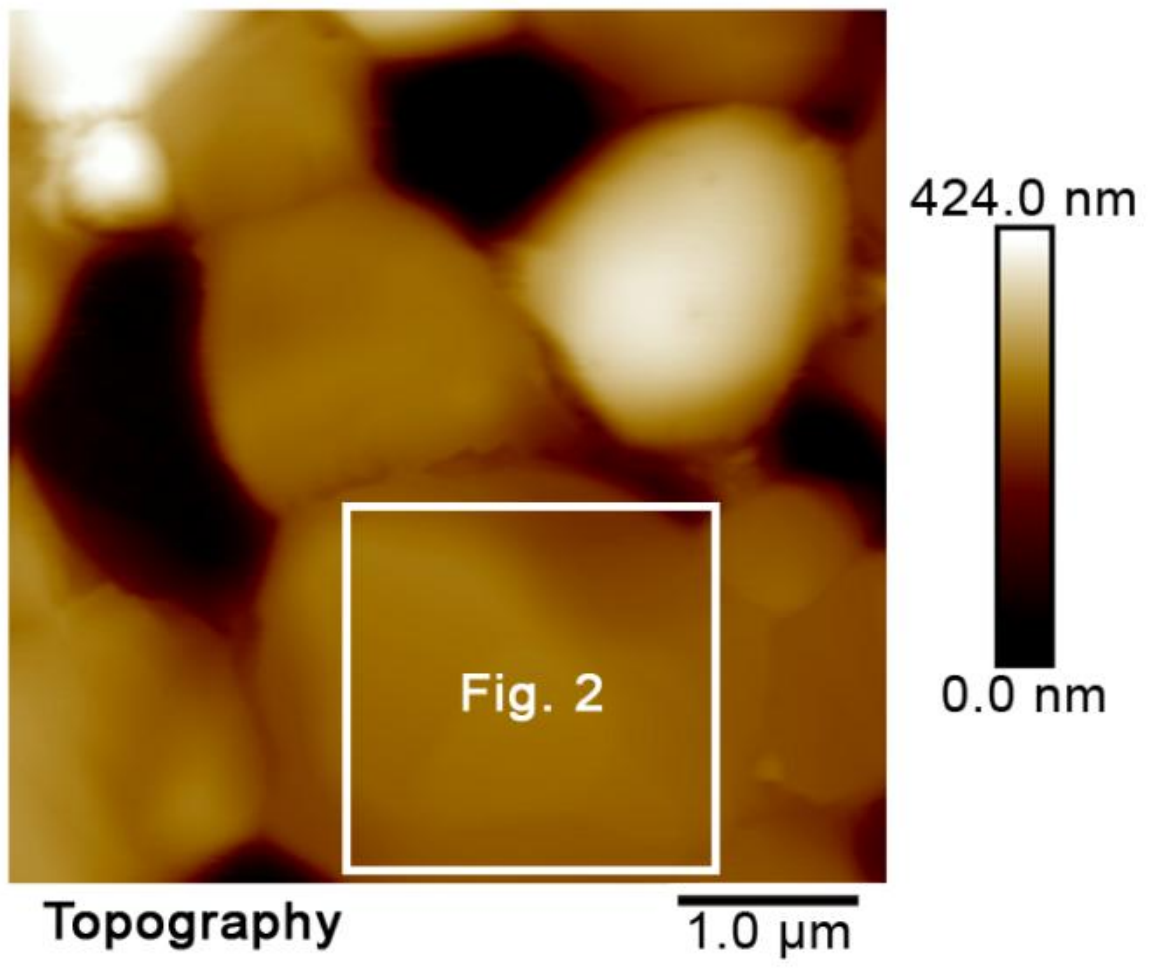

**Fig. S8**. 5 µm × 5 µm AFM topography image showing the environment of the grain depicted in Fig. 2.

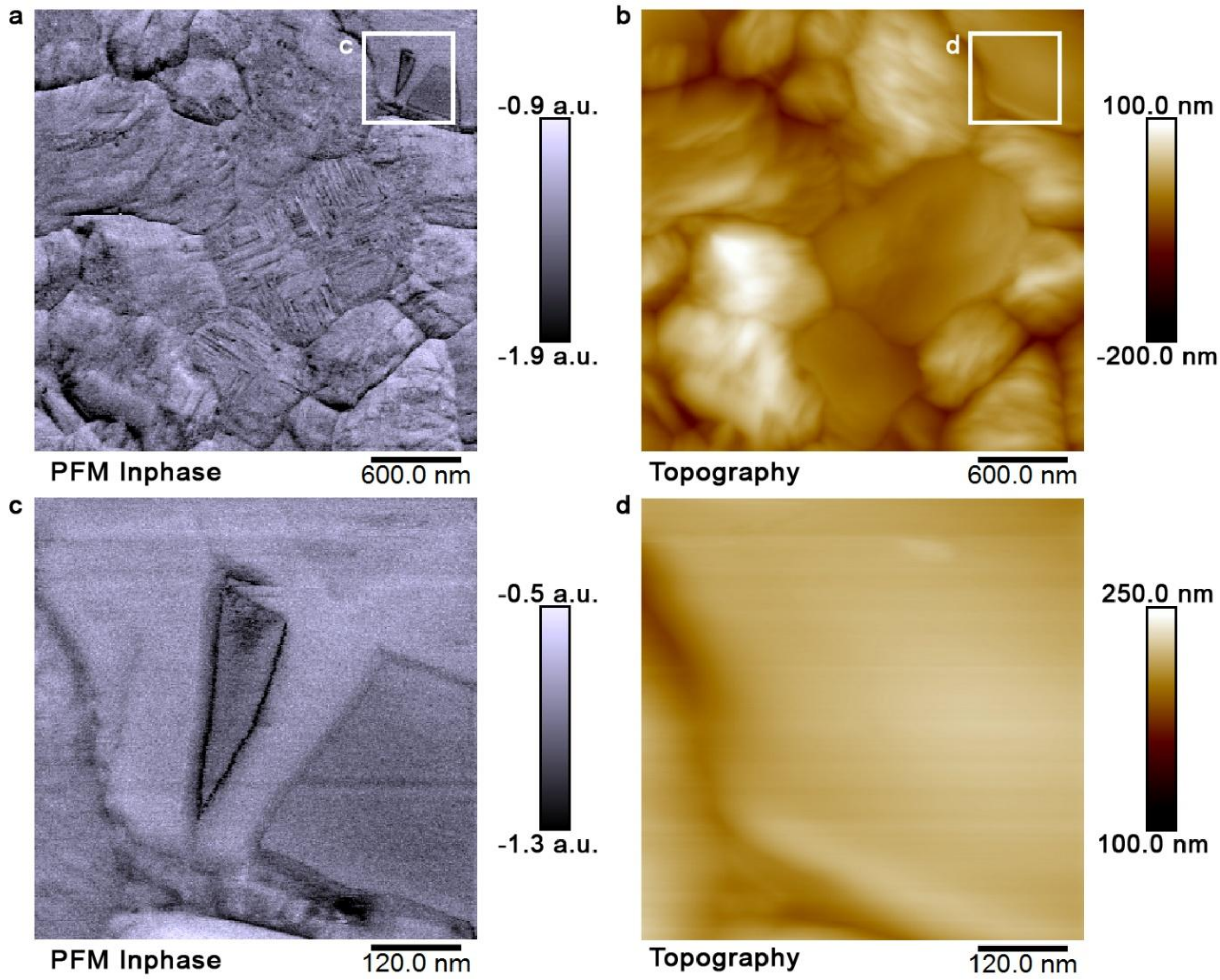

**Fig. S9**. $MAPbI_3$ thin-films solution deposited directly atop an ITO electrode (no PEDOT:PSS layer) omitting methylammonium chloride. Despite the different film forming in absence of methylammonium chloride, the distinct domain features prevail: (a) PFM image: The grains exhibit alternating ferroelectric domains as observed on $MAPbI_3(Cl)$ including various shapes and orientations. (b) The topography image shows a grain surface significantly rougher than $MAPbI_3(Cl)$ samples. (c)+(d) Detailed PFM and topography images show ferroelectric domains that are not visible in the sample topography of a flat grain.

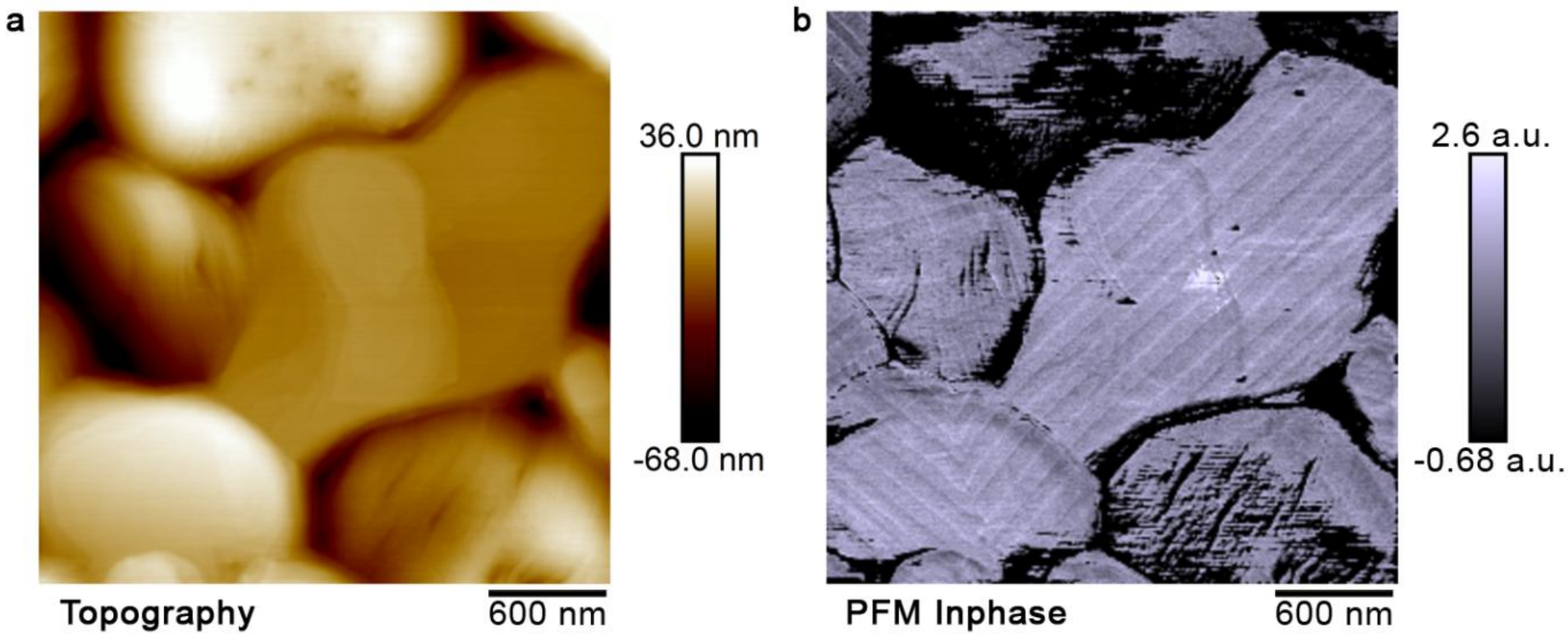

**Fig. S10**. $MAPbI_3$(Cl) grains with a terrace-like surface topography show continuous domains extending over terrace flanks and hence different sample heights. The domains do not show any correlation with the shape of the terraces. While the PFM image exhibits some cross-talk from topography (central area in this image) the domain patterns are clearly distinguishable from topographical features.

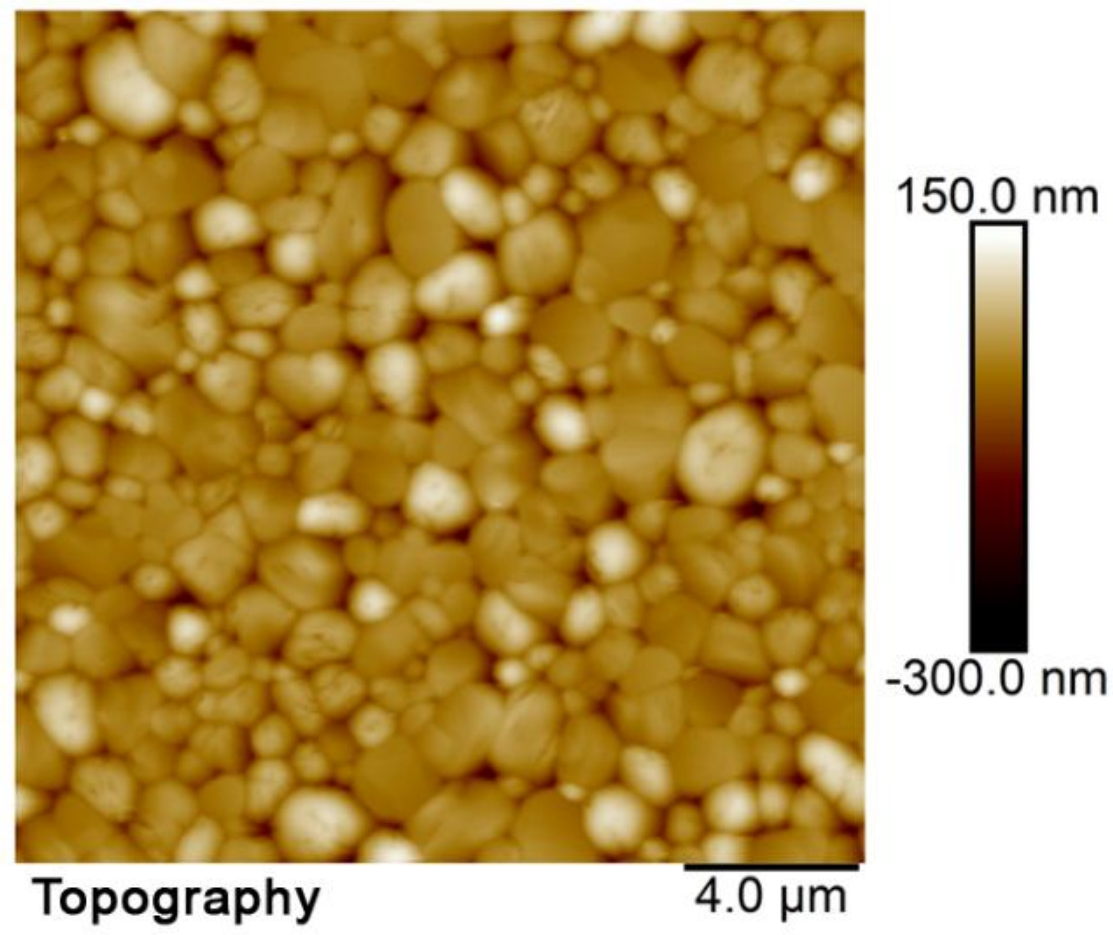

**Fig. S11**. Large area (20 µm × 20 µm) AFM image of a typical sample topography.

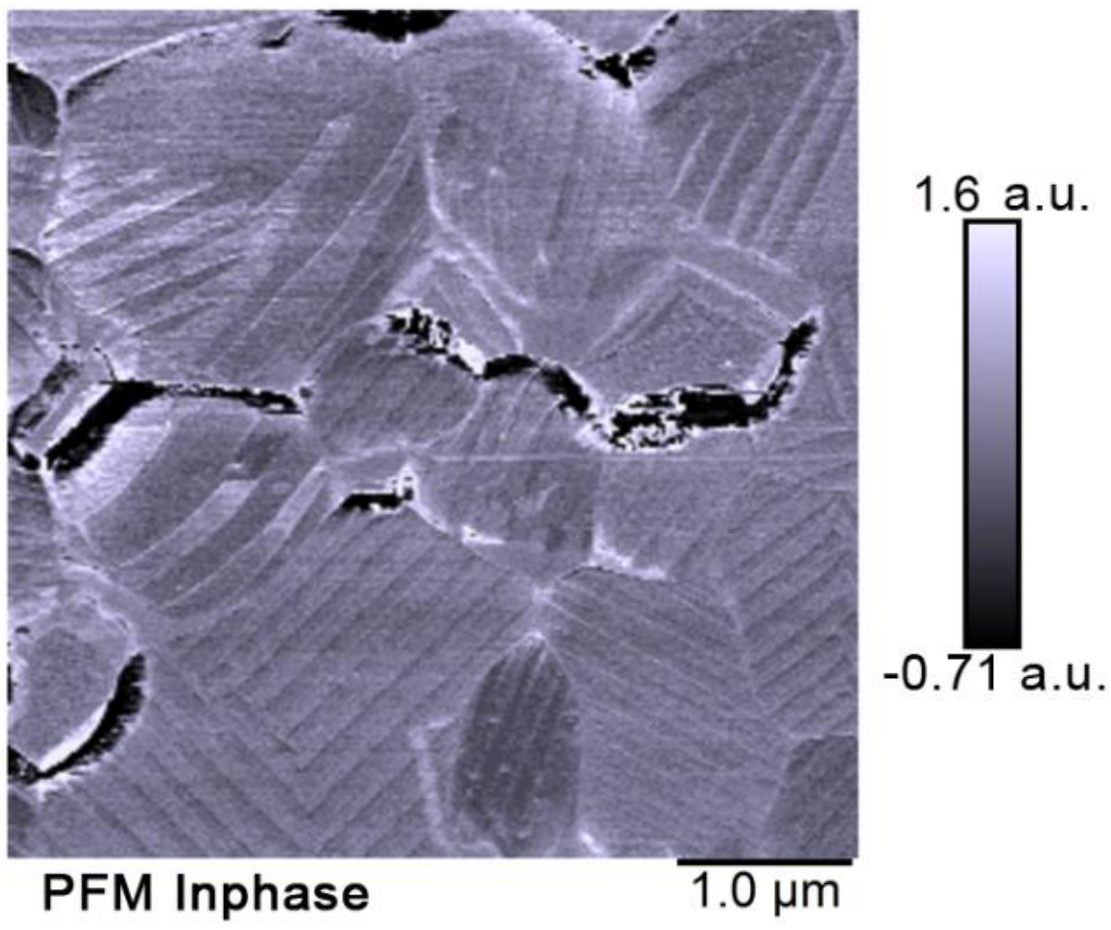

**Fig. S12**. PFM image of a sample region exhibiting rounded and tilted grains resulting in domain patterns of varying width and shape. The flat grains in the bottom half of the image show highly oriented domain patterns.

*Settings for AFM measurements:*
Figure S2a: horizontal, Figure S2b: vertical, 1500 mV AC bias, 34.37 kHz, scan angle 90°, probe SCM-PIC-V2. Figure S3: 1500 mV AC bias, 110.46 kHz, horizontal, scan angle 90°, probe SCM-PIC-V2. Figure S4: Drive amplitude 3000 mV, w/o illumination, probe SCM-PIT. Figure S5: 1500 mV AC bias, 33.14 kHz, vertical, scan angle 90°, probe SCM-PIC-V2. Figure S6a: 1500 mV AC bias, 33.98 kHz, vertical, scan angle 90°, probe SCM-PIC-V2. Figure S6b: under illumination, +800 mV bias, probe SCM-PtSi.
Figure S7a: 1000 mV AC bias, 53.09 kHz, vertical, scan angle 90°, probe SCM-PIC.
Figure S7b: under illumination, w/o bias, probe SCM-PIT. Figure 7c+d: Drive amplitude 3000 mV, w/o illumination, probe SCM-PIT.
Figure S9a+c: 1000 mV AC bias, 41.75 kHz, vertical, scan angle 90°, probe SCM-PIC-V2.
Figure S10b: 1500 mV AC bias, 109.97 kHz, vertical, scan angle 90°, probe SCM-PIC-V2.
Figure 12: 1500 mV AC bias, 62.0 kHz, vertical, scan angle 90° probe SCM-PIC.

Sample fabrication in Figure S9:
$PbI_2$ (Sigma-Aldrich) was dissolved in DMSO (250 mg/ml) and subsequently filtered with a hydrophilic membrane filter (pore size 0.25 µm). Then the solution was blade-coated on a cleaned ITO sample. After 20 s, the films were dried under nitrogen flow to obtain a smooth and homogeneous $PbI_2$ layer. The substrates were transferred into a glovebox with nitrogen atmosphere for spin-coating MAI that was dissolved in IPA (40 mg/ml) without any additives. Further process steps and parameters are identical to the $MAPbI_3$(Cl) sample preparation as described in the Method section of the manuscript.